\documentclass[letterpaper, 10 pt, conference]{ieeeconf}  

\IEEEoverridecommandlockouts                              
\usepackage{amsmath,amssymb,amsfonts}
\usepackage{algorithmic}
\usepackage{graphicx}
\usepackage{xcolor}
\usepackage{booktabs} 
\usepackage{multirow} 
\title{\LARGE \bf
Breast Cancer Classification in Deep Ultraviolet Fluorescence Images Using a Patch-Level Vision Transformer Framework
}
\author{Pouya Afshin$^{1}$, David Helminiak$^{2}$, Tongtong Lu$^{3}$, Tina Yen$^{4}$, Julie M. Jorns$^{5}$, Mollie Patton$^{5}$, Bing Yu$^{3}$, \\Dong Hye Ye$^{1\dag}$ \\ $^1$ Department of Computer Science, Georgia State University\\ $^2$  Department of Electrical and Computer Engineering, Marquette University\\ $^3$ Department of Bioengineering, Marquette University \\ $^4$ Department of Surgery, Medical College of Wisconsin\\ $^5$ Department of Pathology, Medical College of Wisconsin\\ 
\thanks{$^{\dag}$Corresponding Author: \tt\small dongye@gsu.edu}
}

\begin{document}

\maketitle
\thispagestyle{empty}
\pagestyle{empty}

\begin{abstract}
Breast-conserving surgery (BCS) aims to completely remove malignant lesions while maximizing healthy tissue preservation. Intraoperative margin assessment is essential to achieve a balance between thorough cancer resection and tissue conservation. A deep ultraviolet fluorescence scanning microscope (DUV-FSM) enables rapid acquisition of whole surface images (WSIs) for excised tissue, providing contrast between malignant and normal tissues. However, breast cancer classification with DUV WSIs is challenged by high resolutions and complex histopathological features. This study introduces a DUV WSI classification framework using a patch-level vision transformer (ViT) model, capturing local and global features. Grad-CAM++ saliency weighting highlights relevant spatial regions, enhances result interpretability, and improves diagnostic accuracy for benign and malignant tissue classification. A comprehensive 5-fold cross-validation demonstrates the proposed approach significantly outperforms conventional deep learning methods, achieving a classification accuracy of 98.33\%.
\end{abstract}
\section{INTRODUCTION}

Breast cancer continues to be one of the most significant health challenges in the United States, with an estimated 319,750 new cases in 2025, including 316,950 in women and 2,800 in men. Statistics for 2025 also estimate that the disease will cause 42,170 deaths, disproportionately affecting women \cite{b6}. Given that, on average, breast cancer affects approximately one in eight women within the United States, there exists an urgent need for faster, more efficient diagnostic tools, approaches, and treatments \cite{b1, b7, b8}.

Breast-conserving surgery (BCS) or lumpectomy is the preferred treatment option for many patients, as it aims to remove tumors while saving healthy tissue. However, the presence of malignant cells at the surgical margins increases the risk of cancer recurrence. Traditionally, intraoperative margin assessment relies on radiographic imaging and postoperative examination of tissue histology, using hematoxylin and eosin (H\&E) staining. Being a postoperative method, while accurate, a positive margin finding can lead to additional surgeries, causing further significant physical discomfort and emotional distress. Novel technological solutions to reduce these conditions and risks are highly desired \cite{b1, b7, b8}.

Deep ultraviolet fluorescence scanning microscopy (DUV-FSM) has emerged as a groundbreaking technology for dealing with these challenges. Utilizing ultraviolet surface excitation, DUV-FSM enables real-time, high-resolution imaging of entire tissue surfaces without destructive sectioning. This modern technique offers exceptional visual clarity, allowing clear differentiation of textures and colors distinguishing cancerous and healthy tissues, making it particularly effective in delineating tumor margins \cite{b1, b7, b8}. Integration of DUV-FSM imaging with automated breast cancer classification techniques offers improved intraoperative margin assessments and minimization of surgical involvement \cite{b1}.

Convolutional neural networks (CNNs) are widely used for medical image analysis, including breast cancer classification. However, the high resolution of DUV whole surface images (WSIs) requires excessive GPU memory for deep learning models, rendering direct processing impractical. Existing works have explored patch-based strategies, dividing WSIs into smaller, manageable portions. For example, \cite{b1} had ResNet-50 extract features, classifying with XGBoost, and focusing attention using Grad-CAM++ saliency maps, while \cite{b2} trained a support vector machine on patch-level features, applying weighted voting for final classification. 

Despite advancements in CNN-based architectures, they are generally limited by a reliance on small receptive fields, preventing the capture of global contextual information \cite{b25}. Since each patch contains interrelated cellular structures and/or cancerous cells, understanding long-range inter-spatial relationships can improve classification accuracy. This study alternately employs a vision transformer (ViT) \cite{b9}, which excels through the use of self-attention mechanisms to capture local and global structural dependencies, enhancing feature learning and providing resilience to noise \cite{b28}.

\begin{figure*}[ht] 
    \centering
    \includegraphics[width=\textwidth]{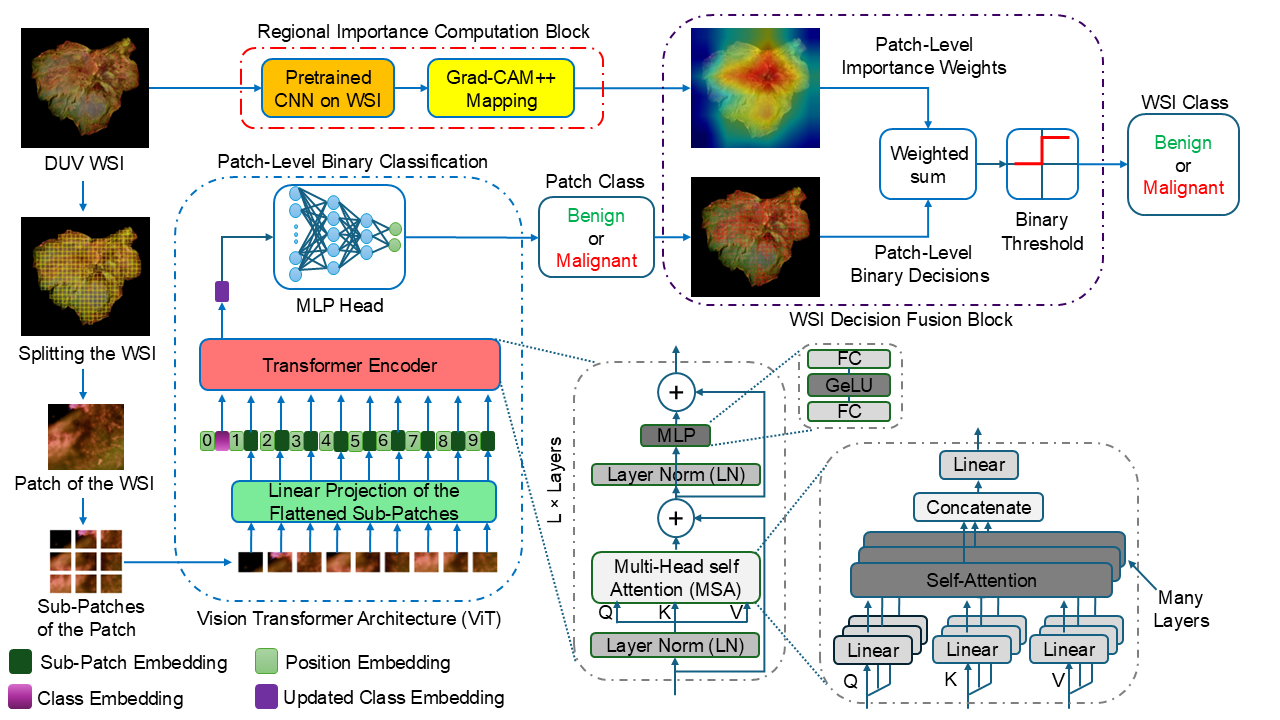} 
    \caption{System Model for DUV WSI Classification: A WSI is divided into non-overlapping patches individually processed by the vision transformer (ViT). Each patch is further divided into sub-patches, transformed into learnable position and class embeddings, and passed through the transformer encoder. The updated class embedding is then classified via the MLP head. Simultaneously, Grad-CAM++ maps, generated using a fine-tuned CNN, provide patch-level importance weights. Finally, patch-level predictions and Grad-CAM++ weights are fused for the WSI-level classification.}
    \label{fig:System Model}
\end{figure*}

In order to improve the interpretability of ViT outputs, Grad-CAM++ \cite{b21} was integrated into a unified decision fusion framework to enhance model transparency and guide WSI-level classification in relevant spatial regions, where CAM \cite{b19} requires retraining per class, and Grad-CAM \cite{b20} assumes equal gradient weights, Grad-CAM++ isolates positive contributions to class predictions. This assigns more accurate per-pixel weights and improves saliency map precision. Guiding patch-level predictions with Grad-CAM++ enhances WSI classification performance.

Summarily, this study makes the following contributions:
\begin{itemize}
\item \textbf{Patch-Based Processing for Deep Learning of WSI} \newline Improves classification performance and efficiency using patch-based processing of high-resolution WSIs. 
\item \textbf{Adoption of ViT Model for Patch Classification} \newline Implements ViT architecture to utilize local and global structural relationships for improved tissue assessments.
\item \textbf{Integration of Grad-CAM++ for Interpretability} \newline Enhances interpretability of ViT results and guides patch-level predictions for improved WSI classification.
\end{itemize}

\section{Method}
This study proposes a new approach for the classification of breast cancer in DUV WSIs. To address the challenge of limited data and enhance tumor localization, each WSI is preprocessed and segmented into smaller, non-overlapping patches. Using a transfer learning strategy, a pre-trained ViT model is fine-tuned to perform patch classification, learning discriminative features from available DUV data \cite{b9}. Following the methodology in \cite{b1}, an approach was adopted to enhance the interpretability of the classifier. Fine-tuning a pre-trained DenseNet-169 \cite{b16} network on DUV WSI data and applying Grad-CAM++ \cite{b21} generates regional saliency maps. These visually represent the model decision-making process and highlight spatial regions important to the classification task. To obtain the WSI-level classification, patch-level predictions are multiplied by their corresponding saliency map scores to compute a weighted combination. Thresholding the results through a sign function then yields a final binary determination.
\subsection{Patch-Level Classification with ViT}
Leveraging transfer learning with a pre-trained ViT model, where common features are already understood, reduces the risk of overfitting and accelerates training \cite{b11}. After fine-tuning with DUV data, the model can extract fine-grained features and classify patches as malignant or benign.

Let \( x_i \) represent the \( i \)-th sample from the DUV WSI dataset, where \( i \in \{1, \dots, M\} \). Partitioning each $x_{i}$ into smaller, non-overlapping patches of \( 400 \times 400 \) pixels allows computational efficiency while providing sufficient detail of tissue morphology and cellular structures to make effective margin assessments. Patches containing more than $80\%$ background are excluded, as outlined in \cite{b1}. Each remaining patch \( \mathbf{p}_i^j \) becomes an individual input, being downsampled to a height \(H\) and width \(W\) required to pass into the ViT-Base architecture (\(224 \times 224\) pixels herein). When subsequently sent through the ViT, each input \( \mathbf{p}_i^j \) gets divided into \( N \) smaller, non-overlapping sub-patches \( \mathbf{s}_{i}^{jk} \in \mathbb{R}^{P \times P \times C} \), where \( k \in \{1, \dots, N\} \) indexes the sub-patches, \( P \times P \) is the sub-patch size, and \( C \) indicates the number of channels. The total number of sub-patches is given by \( N = \frac{H \times W}{P^2} \) \cite{b9}.

Each sub-patch \( \mathbf{s}_i^{jk} \) then gets flattened into a vector \( \text{vec}(\mathbf{s}_i^{jk}) \in \mathbb{R}^{1 \times (P^2 \cdot C)} \) and linearly projected to a fixed dimension \( D \) by means of a trainable projection matrix \( \mathbf{E} \in \mathbb{R}^{(P^2 \cdot C) \times D} \) \cite{b9}. This operation is defined as:
\begin{align}
\hat{\mathbf{s}}_{i}^{jk} = \text{vec}(\mathbf{s}_i^{jk}) \mathbf{E} \in \mathbb{R}^{1 \times D}.
\end{align}
Each \( \hat{\mathbf{s}}_{i}^{jk}\) represents one of \( N \) sub-patch embeddings within an input sequence. Providing an aggregate, global patch representation, a trainable class embedding \( \mathbf{z_0^0} = \mathbf{p}_i^{j \text{class}} \in \mathbb{R}^{1 \times D} \) is concatenated at the start of the sequence \cite{b9}. A trainable positional embedding \( \mathbf{E}_i^{j \text{pos}} \in \mathbb{R}^{(N+1) \times D} \) is then added to the sequence of embeddings, preserving patch-level location information for each sub-patch. The completed input sequence for the transformer encoder is formed by:
\begin{align}
\mathbf{z}_0 = [\mathbf{p}_i^{j \text{class}}, \hat{\mathbf{s}}_{i}^{j1}, \hat{\mathbf{s}}_{i}^{j2}, \dots, \hat{\mathbf{s}}_{i}^{jN}] + \mathbf{E}_i^{j \text{pos}}.
\end{align}
A transformer encoder processes an input \( \mathbf{z}_0 \) through layers \( \ell \in \{1, \dots, L\} \) comprising layer normalization (LN), multi-head self-attention (MSA) and multi-layer perceptron (MLP) blocks (two linear layers with separating GELU activation), both blocks incorporating residual connections \cite{b9}. For each layer, the operations are defined as follows:
\begin{align}
\mathbf{z}^{\prime}
_{\ell} &= \text{MSA}(\text{LN}(\mathbf{z}_{\ell-1})) + \mathbf{z}_{\ell-1}, \\
\mathbf{z}_{\ell} &= \text{MLP}(\text{LN}(\mathbf{z}^{\prime}_{\ell})) + \mathbf{z}^{\prime}_{\ell}.
\end{align}

The resulting class token \( \mathbf{z}_{L}^{0}\) undergoes a last layer normalization and passes into a single linear layer \( \text{MLP}_{head} \) to classify the \( j \)-th patch from the \( i \)-th WSI, \( \mathbf{p}_i^j \) as follows \cite{b9}:
\begin{align}
\hat{y}_i^j &=\text{MLP}_{head}(\text{LN}(\mathbf{z}_{\text{L}}^0)).
\end{align}
The predicted \( \hat{y}_i^j \) and ground-truth label \( y_i^j \) for a patch \( \mathbf{p}_i^j \) use values of 0 and 1 to denote benign and malignant tissues. 

%

\subsection{WSI-Level Fusion with Grad-CAM++}
Let \( Y_{i}^{y_i} \) denote the output logit value for the \( i \)-th WSI sample \( x_i \) and class corresponding to the ground-truth label \( y_i \). The logit represents the raw model output before the application of a softmax function \cite{b21}. Let \( F^q \) represent the \( q \)-th of \( Q \) feature maps of a selected layer, where \( q \in \{1, \dots, Q\} \). The importance weight assigned to a specific feature map \( F^q \) for each class \( y_i \) is defined through: 
\begin{align}
\lambda_{q}^{y_i} &= \sum_{\alpha} \sum_{\beta} w_{\alpha\beta}^{qy_i} \cdot \text{ReLU} \left( \frac{\partial Y_{i}^{y_i}}{\partial F_{\alpha \beta}^{q}} \right),
\end{align}
where the indices \( \alpha \) and \( \beta \) represent a pixel location and the weighting coefficient \( w_{\alpha\beta}^{q y_i} \) denotes its importance (refer to \cite{b21} for derivation). The gradient \( \frac{\partial Y_{i}^{y_i}}{\partial F_{\alpha \beta}^{q}} \) represents the derivative of the raw output score \( Y_{i}^{y_i} \) with respect to the feature map \( F^q \) at position \( (\alpha, \beta) \). The saliency map \( R_{i} \) for the \( i \)-th WSI sample, \( x_i \), is obtained by:
\begin{align}
R_{i}&= \text{ReLU}\left( \sum_{q} \lambda_{q}^{y_i}\cdot F^q \right).
\end{align}
The ReLU function effectively limits considered features and pixels to those with positive contributions to class prediction, emphasizing the most informative and relevant regions \cite{b21}. 

Similar to \cite{b22}, Grad-CAM++ was applied on a pre-trained DenseNet-169 model to create saliency maps using features extracted from the batch normalization layer between the final convolutional and classification layers. Let \( A_{i}^{j} \) denote the region corresponding to the \( j \)-th patch of the \( i \)-th WSI sample, \( x_i \). The patch saliency score averages the saliency map values over the pixel area \( A_{i}^{j} \), being determined through:
\begin{align}
r_{i}^{j} = \frac{1}{|A_{i}^{j}|} \sum_{(\alpha, \beta) \in A_{i}^{j}} R_{i,\alpha \beta}.
\end{align}
Let \( |A_{i}^{j}| \) denote the total number of pixels in the \( j \)-th patch (herein \( 400 \times 400 \) pixels) and the saliency value at position \( (\alpha, \beta) \) be represented by \( R_{i,\alpha \beta} \). Initial ViT model predictions, with labels of 0 (benign) or 1 (malignant), for a patch \( \mathbf{p}_i^{j} \) are remapped to -1 and +1, then denoted as \( \tilde{y}_i^j \). Applying weighted majority voting fuses the transformed values with corresponding saliency scores. An empirical threshold \( r_{i}^j > 0.30 \) ensures that only patches with significant saliency contribute to the weighted sum. This approach weights patch-level contributions to WSI-level classification according to the determined relevance of evaluated features. Resulting non-zero values are mapped via a function \( \text{sign}(\cdot) \), to binary values, where \( \mathbb{R}-\mapsto -1 \) and \(\mathbb{R}+\mapsto +1\) \cite{b22}. Overall, the final predicted label for $x_{i}$ is computed to be:
\begin{align}
\hat{y}_i =
\begin{cases} 
\text{sign} \left( \frac{1}{N_i} \sum_{j=1}^{N_i} \left[ r_{i}^j \cdot \tilde{y}_{i}^j \right] \right) \quad & \text{if } r_{i}^j > 0.30 \\
0 & \text{otherwise}.
\end{cases}
\end{align}

\begin{figure*}[ht] 
    \centering
    \includegraphics[width=\textwidth]{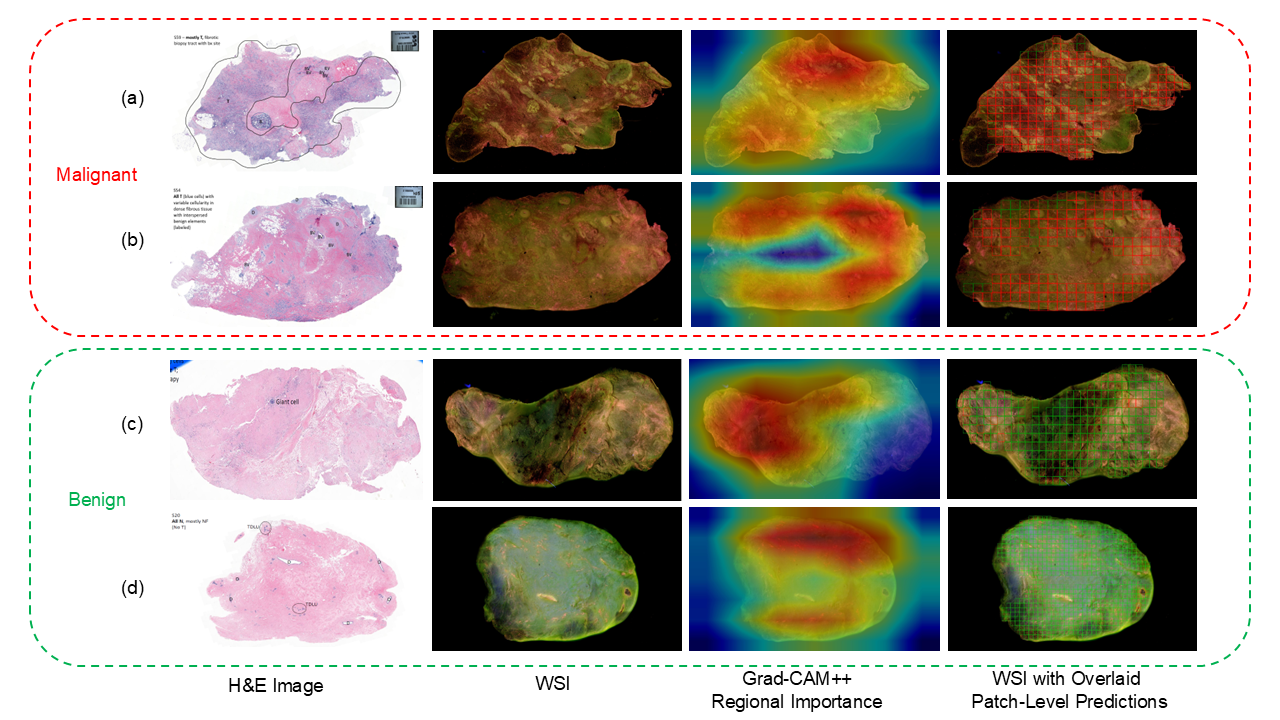} 
    \caption{Visualization of DUV WSIs with their corresponding H\&E images, Grad-CAM++ saliency maps, and Patch-level predictions. Cases (a) and (b) show malignant samples, while (c) and (d) represent benign ones. ViT outperforms CNNs in patch-level predictions, accurately determining patch labels in most cases. While the ViT misclassified some patches in (b) and (c), their impact was mitigated by Grad-CAM++ saliency scores. The proposed method refines WSI classification by heavily weighting diagnostically important regions and de-emphasizing less critical areas.}
    \label{fig:Results}
\end{figure*}

\section{EXPERIMENTS}
This study developed a method to classify DUV WSI collected during breast-conserving surgery (BCS) as benign or malignant. A 5-fold cross-validation was performed to ensure a robust evaluation, with separated training and test sets. Predefined folds of WSI samples were set, with one fold reserved in each iteration for testing, 80\% of the remaining for training, and the other 20\% for validation. 

A pre-trained ViT-Base model (ViT-B/16) was fine-tuned on patches derived from training folds, using stochastic gradient descent (SGD) \cite{b17}, a learning rate of $3 \times 10^{-4}$, and a cosine learning rate scheduler. The model that performed best across the validation sets was selected for final evaluation on the test set. Focusing predictions on relevant regions and enhancing interpretability, Grad-CAM++ was integrated with a pre-trained and fine-tuned Densenet-169 \cite{b16}. 

For comparison and Grad-CAM++ analysis, the cross-validation was repeated using a pre-trained DenseNet-169 network \cite{b16}. This model was fine-tuned with an Adam optimizer \cite{b18} at a learning rate of \( 10^{-4} \) for 30 epochs, using a 40\% dropout rate to match \cite{b22}. Another pre-trained ViT model was similarly independently tuned at a \( 10^{-3} \) learning rate and evaluated. For both, the WSIs were resized to \(224 \times 224 \) pixels to match input dimension requirements.

\begin{table*}[ht]
    \centering
    \caption{WSI Classification Performance with 5-Fold Cross-Validation [Unit: \%] }
    \label{tab:performance}
    \renewcommand{\arraystretch}{1.2} 
    \begin{tabular}{|c|c|c|c|c|c|c|}
        \hline
        \multirow{2}{*}{\textbf{Method}} & \multirow{2}{*}{\textbf{Decision Fusion}} & \multicolumn{5}{c|}{\textbf{Performance Metrics}} \\
        \cline{3-7} 
         &  & \textbf{Accuracy} & \textbf{Precision} & \textbf{F1 Score} & \textbf{Sensitivity} & \textbf{Specificity} \\ 
        \hline
        ResNet-50* \cite{b1} & N/A & 81.67 & 80.49 & 85.71 & 91.67 & 66.67 \\ 
        \hline
        DenseNet-169 \cite{b16} & N/A & 86.67 & 88.89 & 88.89 & 88.89 & 83.33 \\ 
        \hline
        ViT \cite{b9}  & N/A & 86.67 & 88.89 & 88.89 & 88.89 & 83.33 \\ 
        \hline
        ResNet-50 + XGBoost* \cite{b1}  & Majority Voting & 93.33 & 94.44 & 94.44 & 94.44 & 91.67 \\ 
        \hline
        ResNet-50 + XGBoost* \cite{b1}  & Grad-CAM++ & 95.00 & 92.31 & 96.00 & 100.00 & 87.50 \\ 
        \hline
        {PatchViT }
         & \textbf{Grad-CAM++} & \textbf{98.33} & \textbf{97.00} & \textbf{99.00} & \textbf{100.00} & \textbf{96.00} \\ 
        \hline
    \end{tabular}
    \vspace{0.2cm} 
    \begin{flushleft}
   \center \footnotesize *Implemented by the authors in \cite{b1} and included for comparison purposes.
    \end{flushleft}
\end{table*}
\subsection{Dataset}
For training and evaluation, a DUV WSI dataset for breast cancer analysis, consisting of 60 tissue samples (24 benign and 36 malignant), was sourced from the tissue bank at the Medical College of Wisconsin \cite{b14}. Therefrom, 34,468 image patches (with \( 400 \times 400 \) pixels) were extracted, including 9,444 malignant and 25,024 benign instances, with ground-truth labels assigned according to pathologist annotations \cite{b1}. 
\subsection{Qualitative Evaluation}
Herein are qualitative evaluations of the breast cancer detection results on malignant and benign WSIs. Figure 2 illustrates examples from both class categories, including the corresponding H\&E images with pathologist annotations, WSIs, and Grad-CAM++ saliency maps. Patch-level predictions are overlaid on the WSIs, with low-importance regions removed by an empirical threshold to emphasize the most diagnostically relevant areas. Grad-CAM++ highlights class-relevant regions with a more intense reddish hue, indicating areas where the fine-tuned DenseNet-169 model focused during classification. This decision fusion weighting mechanism was observed to prevent incorrect patch-level predictions made by the ViT, as evident in cases (b) and (c). For case (b), there were tumor cells characterized by variable cellularity within dense fibrous tissue and interspersed benign elements. Case (c) presented predominantly normal tissue with scattered small inflammatory blue cells and no definite tumor. In both cases, these misclassified patches were de-emphasized by the Grad-CAM++ weighting, correcting the WSI-level classifications.

\subsection{Quantitative Evaluation}
Table I compares cross-validation classification performance, with matched folds across the DUV WSI dataset, between prior \cite{b1} and proposed methods. ViT, DenseNet-169, and ResNet-50 \cite{b30} all demonstrated relatively poor performance. The internal downsampling operations of the CNN models lose critical contextual information. Restriction of fine-grained details of cellular structures, cancerous regions, and other essential components at the early stages of these architectures limited their ability to accurately distinguish benign and malignant samples. The large model sizes and relatively small dataset exacerbated overfitting issues, further degrading performance. These behaviors matched expectations since CNNs are known to struggle to capture global spatial contexts, and ViTs typically require copious quantities of data to achieve optimal performance \cite{b25}, \cite{b28}.

In contrast, the ViT model implemented by this work processes WSI at a patch-level. Leveraging self-attention to capture local and global contextual details enables more precise, patch-level assessments. The results were then refined by fusing classifier decisions with Grad-CAM++ saliency maps, particularly effective in cases where the ViT alone struggled or made errors due to a lack of relevant features. Grad-CAM++ weighting emphasized critical regions, suppressing irrelevant or low-confidence areas. This approach achieved an impressive $98.33\%$ accuracy, surpassing other examined deep learning methods up to approximately $13\%$. 

The fusion method was compared with the approach in \cite{b1}, which utilized ResNet-50 for feature extraction, XGBoost for classification, and Grad-CAM++ for regional importance weighting. Despite the improvement over traditional deep learning approaches, the CNN struggled to capture long-distance spatial relationships. Limited receptive fields reduced the extraction of meaningful features needed to distinguish cancerous and benign tissues. As a result, the updated approach achieved a 3\% higher relative accuracy.

Analysis of other performance metrics was conducted, including precision: the proportion of true positive predictions; sensitivity: the proportion of actual positives correctly identified; and specificity: the proportion of actual negatives correctly identified. A limitation evidenced by the prior CNN formulation was a tendency to overpredict malignancy. Modernizing to a ViT architecture has achieved gains of about 5\% and 9\% in precision and specificity, while maintaining perfect recall. Practically, this demonstrates improved network detection of benign cases and reduces the number of false positives during WSI classification.

Overall, the new model's improved precision and specificity make it a more reliable tool for medical imaging. The approach offers significant clinical benefits, where patch-level assessment may provide surgeons with precise localization of cancerous material during operations. The Grad-Cam++ saliency maps additionally indicate model confidence regarding predictions, reducing the risk of unnecessary actions on incorrect positive margin-level detections or potential determination of the need for re-excision surgeries.

\section{Conclusion}
This study presents an automated method for classifying DUV WSIs to differentiate benign and malignant tissues during breast-conserving surgery (BCS) and accurately localize cancerous cells. Given the computational limitations of deep learning models when processing high-resolution WSIs, a patch-level framework was adopted to improve computational tractability and efficiency. Through a robust 5-fold cross-validation, it was demonstrated that vision transformer (ViT) model performance may be enhanced with Grad-CAM++ weighted decision fusion. The ViT effectively extracts both local and global features from WSI patches, while Grad-CAM++ refines patch-level predictions, leading to more accurate WSI classification, even in particularly challenging cases. The proposed method should provide valuable insights to surgeons and pathologists, reducing the risk of incorrect cancer margin detection and minimizing the need for additional surgeries. 

\end{document}